\documentclass[aps,prl,amsmath,amssymb,reprint,superscriptaddress]{revtex4-2}

\usepackage{graphicx}
\usepackage{textcomp}
\usepackage{xcolor}

\usepackage{soul}

\newcommand{\sst}[1]{\scriptscriptstyle{#1}}
\newcommand{\LL}{ {\sst{L}} } 
\newcommand{\HH}{ {\sst{H}} } 
\newcommand{\DP}{ \mathrm{\sst{DP}} }
\newcommand{\DO}{ \mathrm{\sst{DO}} }
\begin{document}

\title{Reversible Trapping of Colloids in Microgrooved Channels\\ via
Diffusiophoresis under
Steady-State Solute Gradients}

\author{Naval Singh}
\author{Goran T. Vladisavljevi\'{c}}
\affiliation{
	Department of Chemical Engineering, Loughborough University, Loughborough, LE11 3TU, United Kingdom}
\author{Fran\c{c}ois Nadal}
\affiliation{
Wolfson  School  of  Mechanical,  Electrical  and  Manufacturing  Engineering, 
Loughborough  University, Loughborough,  LE11  3TU,  United  Kingdom}
\author{C\'{e}cile Cottin-Bizonne}
\author{Christophe Pirat}
\affiliation{
Institut Lumi\`{e}re Mati\`{e}re, UMR5306 Universit\'{e} Claude Bernard Lyon 1
- CNRS, Universit\'{e} de Lyon, Villeurbanne Cedex, 69622, France} 
\author{Guido Bolognesi}
\email{g.bolognesi@lboro.ac.uk}
\homepage{www.particlemicrofluidics.com}
\affiliation{
	Department of Chemical Engineering, Loughborough University, Loughborough, LE11 3TU, United Kingdom}
\date{\today}

\begin{abstract}
The controlled transport of colloids in dead-end structures is a key capability that can
	enable a wide range of applications, such as
	bio-chemical analysis, drug delivery and underground oil recovery.
	This letter presents a new trapping mechanism that
	allows the fast (i.e., within a few minutes) and reversible accumulation of
	sub-micron particles within dead-end micro-grooves by means of parallel
	streams with different salinity level. For the first time, particle focusing
	in dead-end structures is achieved under steady-state gradients.
	Confocal microscopy analysis and
	numerical investigations show that the particles are trapped at a
	flow recirculation region within the grooves due to a combination of diffusiophoresis
	transport and hydrodynamic effects.  Counterintuitively, the 
	particle velocity at the focusing point is not vanishing and, hence, the
	particles are continuously transported in and out of the focusing point.
	The accumulation process is also reversible and one
	can cyclically trap and release the colloids by controlling the salt concentration
	of the streams via a flow switching valve.
\end{abstract}

\maketitle

Particle transport in confined structures plays an important role in several technological applications,
including drug delivery, diagnostics, enhanced oil recovery, particle separation and
filtration technologies.
Nevertheless, the implementation of an effective strategy for controlling the motion of
colloidal particles within a confined environment, such as a dead-end channel
or a porous medium, is still a challenging and thought-provoking task. 
In recent years, an increasing
number of studies have exploited the motion of particles and liquids induced by
solute concentration gradients --- the so-called diffusiophoresis (DP) and
diffusioosmosis (DO) phenomena --- to enable particle manipulation capabilities,
such as delivery to/extraction from dead-end
pores~\cite{kar2015enhanced,shin2016size,shin2018cleaning}, 
particle focusing~\cite{Abecassis2008d,Palacci2010b,Shi2016b,shin2017accumulation,friedrich2017molecular}
and separation~\cite{Shin2017d,ha2019dynamic,rasmussen2020size}. 
In DP with electrolytes, the motion of a particle is driven by a solute
concentration gradient $\mathbf{\nabla} c$ and the resulting particle velocity can be
expressed as $\mathbf{u}_{\DP}=\Gamma_{\DP} \mathbf{\nabla} \ln c$, with
the DP coefficient $\Gamma_{\DP}$ being a function of the particle and
solution properties~\cite{Prieve1984b}.
Externally applied unsteady solute gradients have been adopted to boost the
otherwise slow and diffusion-limited migration of nano-/micro-particles within
dead-end structures~\cite{kar2015enhanced,shin2016size,shin2018cleaning}.
However, due to the transient nature of the imposed gradients, the particle and
flow manipulation capabilities are lost within a short period of time
(typically, few tens of minutes).  The ability to generate a steady-state
solute gradient within dead-end structures and, hence, retain indefinitely the
particle manipulation capability has yet to be achieved.  On the other hand,
steady-state solute gradients have been used to accumulate colloids in target
locations within microfluidic chambers and open-ended microchannels.  For instance,
steady-state gradients of chemically reactive solutes~\cite{Shi2016b} can
induce particle focusing at the location where $\mathbf{u}_\DP=0$.
Alternatively, particle focusing can be achieved also by counteracting the DP
particle migration with a hydrodynamic flow, $\mathbf{u}$, that advects
the particles in the opposite
direction~\cite{shin2017accumulation,friedrich2017molecular,rasmussen2020size}. As a result,
particle accumulation occurs at the regions where the particle's total
velocity, $\mathbf{u}_\mathrm{p}=\mathbf{u}+\mathbf{u}_{\DP}$, vanishes.
For sub-micron particles, however, their slow Brownian diffusion cannot
compete with the particle velocity $\mathbf{u}_\mathrm{p}$ and, thus, the particle
concentration increases indefinitely until the packing limit is reached and the
microchannels are irreversibly clogged~\cite{shin2017accumulation}.

Here, we report a new focusing mechanism through which
sub-micron particles can be rapidly and reversibly 
accumulated within dead-end structures (microgrooves) 
by means of a steady-state gradient. In contrast with other focusing strategies,
a steady-state particle distribution is achieved within a few minutes meanwhile the
concentration peak remains well below the packing limit, thereby avoiding
irreversible effects such as particle clustering and device clogging. 
The particle accumulation process is hence fully reversible and colloids
can be transported into and out of the grooves multiple times by 
switching between different flow streams.
%
%
 \begin{figure*}[t!] \includegraphics[width=\textwidth]{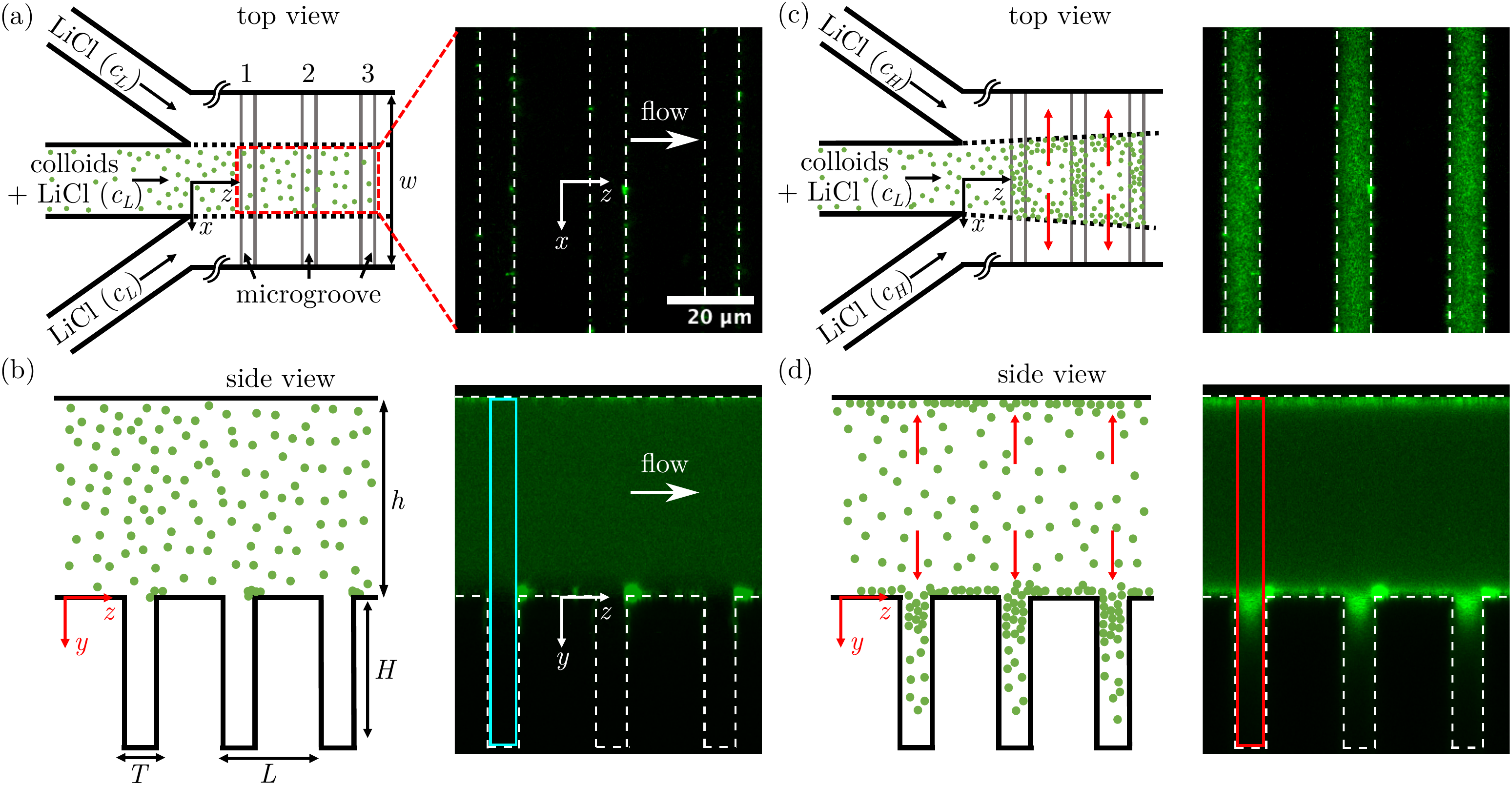}
	 \caption{Solute-induced particle trapping in microgrooves under
	 steady-state conditions. (a, b) Schematics of top and side view of the
	 device and corresponding fluorescence micrographs of three grooves
	 located at $4$~mm from the junction \emph{without} solute contrast.
	 Outer flow: LiCl in DI water at low concentration, $c_\LL=0.1$~mM.
Inner flow: colloids ($215$~nm  in diameter
@~$0.025\%$~v/v) + LiCl in DI water at low
concentration $c_\LL$. (c, d) Schematics and micrographs as in panels (a) and
(b) \emph{with} solute contrast.  Outer flow: LiCl in DI water at high
concentration, $c_\HH=10$~mM. Inner flow: colloids + LiCl in DI water at low
concentration, $c_\LL$.  Red arrows show the direction of the salt gradient.
White dashed lines represent channel boundaries and groove edges.  Channel
size: width $w=400$~{\textmu}m, depth $h=57$~{\textmu}m.  Groove size:
thickness $T=8$~{\textmu}m, depth $H=45$~{\textmu}m, pitch $L=32$~{\textmu}m.
In (a) and (c), fluorescence intensities are averaged along the channel depth
direction $y$, whereas in (b) and (d) intensities are averaged over the
width direction $x$.  The same color scale applies to all
micrographs. Blue and red rectangles show the integration windows over which
the particle concentration profiles in Fig.\ref{fig:figure-2} are calculated.
\label{fig:figure-1}} \end{figure*}

To create a steady-state solute gradient, parallel flows are injected into a
$\Psi$-shaped microchannel, made of an optical adhesive (NOA81) glued on top of
a silicon substrate with transverse microgrooves, as schematically shown in
Fig.~\ref{fig:figure-1}(a) --- see Supplementary Information (SI) for details on
the device fabrication~\cite{SI}.  A total of 1250~grooves are evenly
distributed along the $4$~cm length of the device.  The inner flow is a
suspension of carboxylate polystyrene fluorescent colloids
(Fluoresbrite$^\text{\textregistered}$ YG, 0.20~{\textmu}m, Polysciences) 
at a concentration of $n_0=0.025\%$~v/v, dispersed in a
water (Ultrapure Milli-Q) solution of LiCl (Acros Organics, 99\%) at low
concentration, $c_\LL=0.1$~mM. The outer flow is a LiCl solution at either low
concentration $c_\LL$ (Fig.~\ref{fig:figure-1}a-b) or high concentration
$c_\HH=10$~mM (Fig.~\ref{fig:figure-1}c-d).  Both the inner and outer flow
rates are equal to $12.5$~{\textmu}L/min, resulting in an average speed $U_0$
of $18.3$~mm/s.  The 3D distribution of particles in the channel and the
grooves is measured via laser scanning confocal microscopy as detailed in SI.
In the absence of salt contrast, the colloidal particles hardly penetrate the
grooves likely due to steric and electrokinetic wall-exclusion
effects~\cite{shin2016size,wu1996electrokinetic}, which keep most of the
particles away from the grooved substrate (Fig.~\ref{fig:figure-1}b).  In
presence of a salt contrast, the particles are expected to migrate towards
region at a higher salt concentration, since the DP coefficient of the
particles, $\Gamma_{\DP}$, is positive (see SI for the evaluation of
$\Gamma_{\DP}$).  A higher salt concentration in the outer flow streams
generates a solute gradient along the channel width direction ($x$ axis) --- see
red arrows in Fig.~\ref{fig:figure-1}(c) --- which leads to colloid spreading
along the same direction, as also previously reported in a similar flow
configuration~\cite{Abecassis2008d}. More interestingly, salt gradients arise
also along the depth direction ($y$~axis) --- see red arrows in
Fig.~\ref{fig:figure-1}(d) --- thereby dragging the particles towards the
channel's top flat wall and inside the grooves.  The $y$-component of the salt
gradient is originated by the Poiseuille-like velocity profile in the
rectangular channel.  Indeed at distances $z$ from the junction much smaller
than $U_0h^2/D_\mathrm{s}\simeq4$~cm --- with $D_\mathrm{s}$ the salt diffusivity
--- the salt diffusion process is affected by the non-uniform velocity profile
along the channel depth, and the width of the salt diffusive zone at the
interface between the inner and outer flows decreases with the distance from
the top and bottom walls~\cite{ismagilov2000experimental}.  Consequently, in
the inner region of the channel (i.e., $|x|/w<0.5$) a solute gradient directed
from the bulk towards the walls is established.  It is worth noting that the DP
migration of charged particles towards the channel walls in a parallel flow
configuration has never been reported before and it could be exploited as a
charge-based particle focusing/filtration strategy in microdevices with flat
walls only. Driven by DP, the particles migrate towards the grooves and
accumulate at the groove's entrance as shown in Fig.~\ref{fig:figure-1}(c-d).

 \begin{figure}[t!] \centering
	\includegraphics[width=\columnwidth]{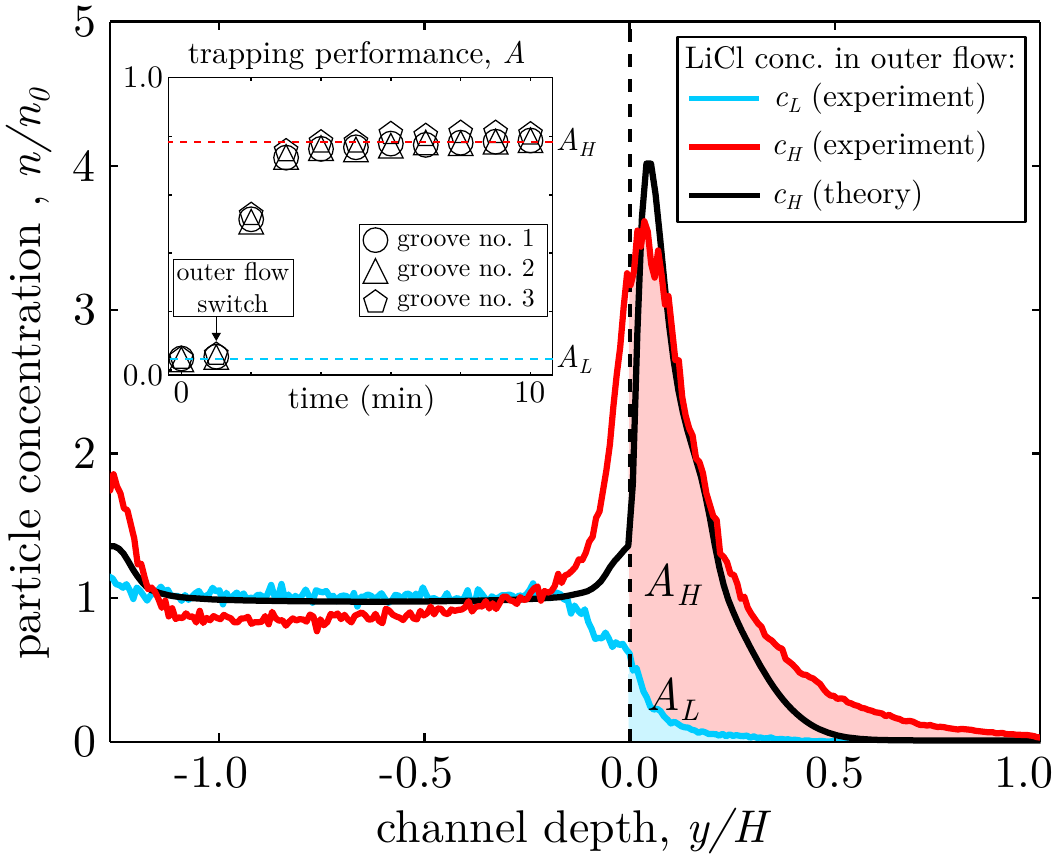}%
	 \caption{
	 Experimental steady-state particle concentration profiles 
along the channel depth \emph{without} salt gradient (blue curve) and
\emph{with} salt gradient (red curve), corresponding
	 to the blue and red integration windows
in Fig.\ref{fig:figure-1}(b) and Fig.\ref{fig:figure-1}(d), respectively.
	 A$_{L}$ and A$_{H}$ are the integral of the profiles
within the groove ($y\ge0$) without and with salt gradient, respectively.
 	The profile (black curve), predicted by the
	numerical simulation in presence of salt gradient, is also shown. 
	Inset: trapping performance of three neighboring grooves, at 4~mm from the junction,
	 as outer flow is switched from low ($c_\LL=0.1$~mM) to 
	 high ($c_\HH=10$~mM) salt concentration.
	\label{fig:figure-2}} \end{figure}

Fig.~\ref{fig:figure-2} shows the steady-state colloid concentration
profiles along the depth direction, with salt gradient (red line) and without
salt gradient (blue line), for groove~1 in Fig.~\ref{fig:figure-1}.  The 3D
colloid concentration field $n(x,y,z)$ are calculated from the fluorescence
intensity of the confocal scan images via a calibration curve (see Fig.~S2).
The concentration profiles in Fig.~\ref{fig:figure-2} are calculated for
each groove by averaging $n(x,y,z)$ over the $x$ range of the confocal images
(ca. $x/w\in[-0.2,0.2]$) and over the $z$ range corresponding to the groove
thickness $T$, as highlighted by the solid rectangles in the side-view
micrographs shown in Fig.~\ref{fig:figure-1}(b,d).  
As shown in Fig.~\ref{fig:figure-2}, the
salt contrast between the parallel streams 
induces the particle migration from the channel bulk towards the
top flat wall ($y\to -h$) and the groove ($y>0$), whereas a slight decrease of
particle concentration in the bulk ($n<n_0$) is observed.  By definition, the
area $A$ below the profile curves for $y>0$ (shaded regions) corresponds to the
average particle concentration within the groove, normalized with respect to
$n_0$.  The parameter $A$ can be hence used as a measure of the groove trapping
performance. The inset in Fig.~\ref{fig:figure-2} shows the evolution of the trapping
performance, $A$, for the three
consecutive grooves shown in Fig.~\ref{fig:figure-1}, as the outer flow is
switched, at the arbitrary time $t=1\:$min, from 
low ($c_\LL$) to high ($c_\HH$) salt concentration solution by means of a flow
switching valve.  In few minutes, the value of $A$ increases from
$A_L=0.022\pm0.002$ to $A_H=0.79\pm0.03$ at steady-state, thereby resulting in
ca. 36 fold increase in the average particle concentration within the grooves.

\begin{figure}[t!] \centering
	\includegraphics[width=\columnwidth]{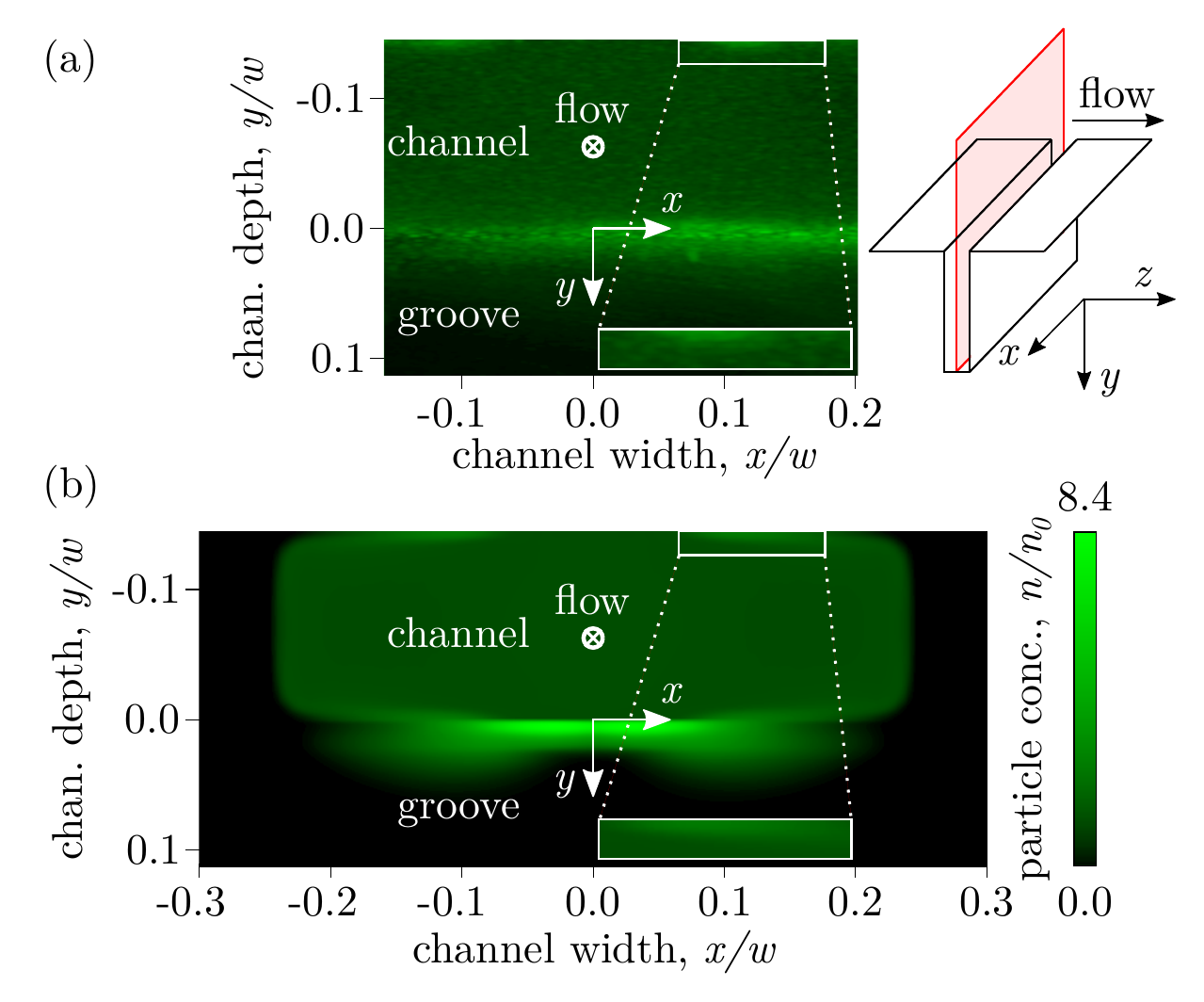}%
	\caption{Experimental (a) and simulated (b) steady-state particle distribution
	on a plane perpendicular to the flow direction, corresponding to the red-shaded region in 
the cartoon, at $4$~mm from the junction and with a solute gradient ($c_\LL=0.1$~mM and $c_\HH=10$~mM). 
	Fluorescence intensities (a) and simulated particle concentration
(b) are averaged over the groove thickness $T$ (along the $z$ axis). The
close-ups highlight one of the two focusing regions near the top flat wall.}
\label{fig:figure-3} \end{figure}
\begin{figure}[t!] \centering
\includegraphics[width=0.9\columnwidth]{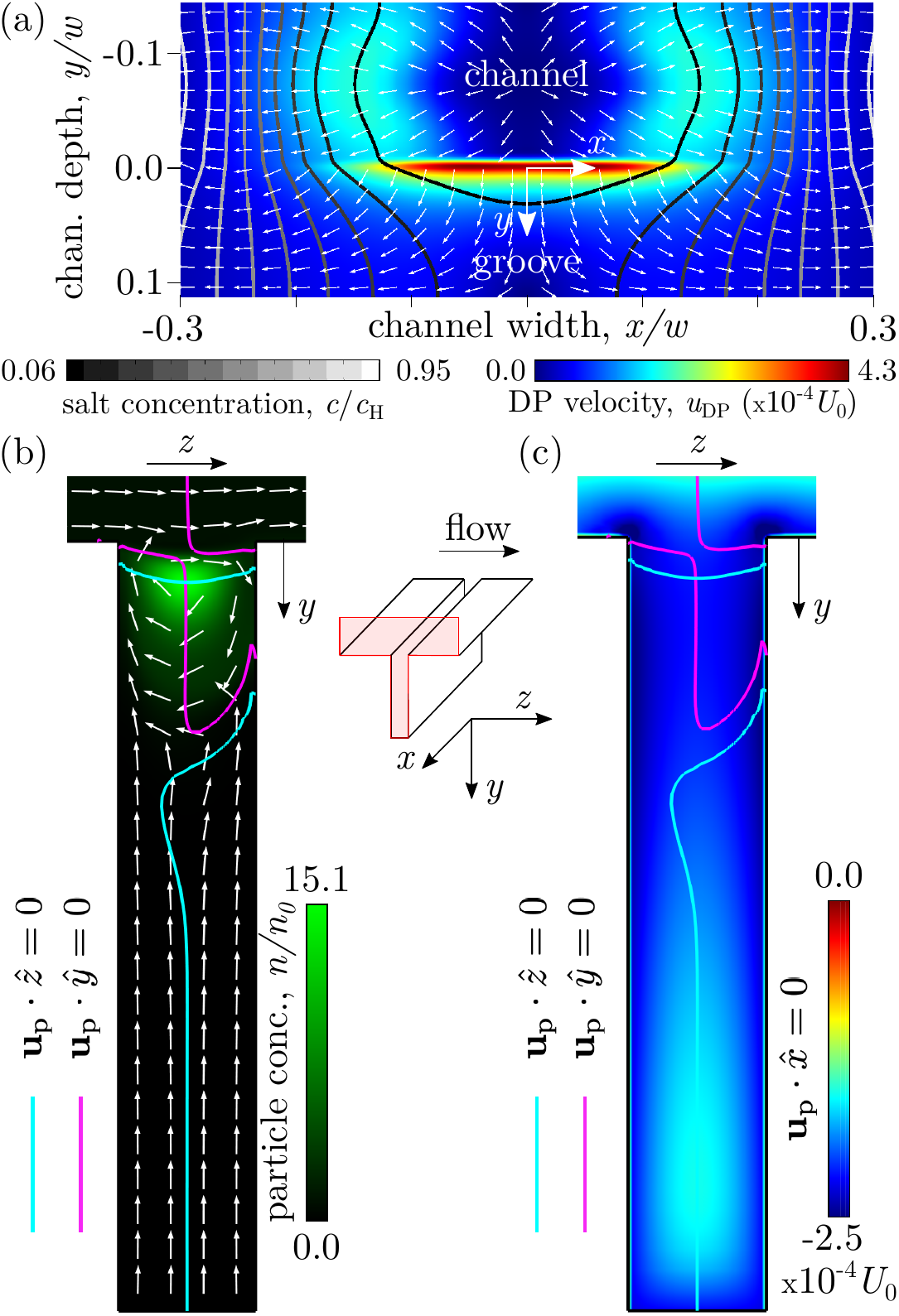}
\caption{Numerical simulation results.
(a) DP velocity intensity and streamlines, and salt
concentration isolines at the same $x$-$y$ cross section as in
Fig.~\ref{fig:figure-3}. (b) Particle concentration and streamlines of the
particle velocity $\mathbf{u}_\mathrm{p}$ at $y$-$z$ cross section, $x/w=0.03$,
corresponding to the red-shaded region in the cartoon. (c) Out-of-plane
component of $\mathbf{u}_\mathrm{p}$ at the same cross section of panel (b).
The blue (magenta) solid lines show the points where the $z$-component
($y$-component) of $\mathbf{u}_\mathrm{p}$ vanishes.  } \label{fig:figure-4}
\end{figure}
To shade light on the mechanisms governing the particle trapping, a numerical
analysis (see SI for details) is performed in Comsol
Multiphysics$^\mathrm{\textregistered}$ to simulate the particle concentration
field.  The 3D computational domain consists of a rectangular channel and a
single groove at $4\:$mm from the junction.  A slip velocity,
$\mathbf{u}_s=-\Gamma_\DO \mathbf{\nabla} \ln c$, with $\Gamma_\DO$ the
diffusioosmosis coefficient, is imposed at the domain walls whereas the
velocity of the particles $\mathbf{u}_\mathrm{p}$ is defined as the sum of the
hydrodynamic velocity and DP velocity,
$\mathbf{u}_\mathrm{p}=\mathbf{u}+\mathbf{u}_\DP$.  The value of $\Gamma_{\DO}$
for the channel walls could not be measured so this parameter is adjusted in
order to achieve a good match between experimental and numerical results (see
SI).  The colloid concentration profile along the channel depth  predicted by
the numerical simulation --- shown as a black curve in Fig.\ref{fig:figure-2}
--- compares well with the experimental profile (red curve).  The good
agreement between experiments and theoretical predictions is confirmed also by
the experimental and simulated particle concentration field on a plane
perpendicular to the flow direction, shown in Fig.\ref{fig:figure-3} (a) and
(b), respectively.  These cross-section views show that the salt gradient leads
to the accumulation of particles just below the groove entrance as well as the
formation of two symmetric and weaker focusing regions, nearly $0.2\:w$
($\simeq 80$~{\textmu}m) apart from each other, close the top flat wall (see
insets). Indeed, in the region $|x|/w<0.5$, the particles are transported
towards the outer flow as well as the top and bottom walls by the DP velocity
field $\mathbf{u}_\DP$, shown in Fig.~\ref{fig:figure-4}(a).  As anticipated,
in the channel the salt concentration isolines, also shown in
Fig.~\ref{fig:figure-4}(a), are bent towards the outer flow due to the
Poiseuille-like hydrodynamic velocity profile.  To clarify why particles
accumulate only at the groove entrance without traveling further deep, one
should look at the particle distribution together with the streamlines of the
velocity field $\mathbf{u}_\mathrm{p}$ at a cross section perpendicular to the
channel widthwise direction.  Fig.~\ref{fig:figure-4}(b) shows such a plot for
the $y$-$z$ cross section, $x/w=0.03$, where the maximum particle concentration
is achieved ($n_{\scriptscriptstyle\max}/n_0\simeq15$).  It can be seen that
the hydrodynamic field $\mathbf{u}$ is characterized by a recirculation region
at the groove entrance and a DO-induced flow with opposite direction with
respect to the particle DP velocity, i.e. outwards of the groove.  Therefore,
as the particles migrate towards the groove by DP, they are captured by the
closed flow streamlines in the recirculation region and accumulate at the
center of the recirculation pattern where the in-plane ($y$ and $z$) components
of $\mathbf{u}_\mathrm{p}$ vanish.  However, the DP migration further down the
groove is counteracted by the DO flow in the opposite direction. 
Importantly, in the absence of DO, the particles would still accumulate within
the recirculation
region, but they would also concentrate at the bottom end of the groove due to
DP transport (Fig. S4). It can be concluded that the observed particle trapping
is due to the combined effects of DP particle migration and hydrodynamic flow
recirculation within the groove. Despite both DO and Brownian diffusion affects
the intensity of the particle concentration peak (i.e.,
$n_{\scriptscriptstyle\max}/n_0$), they are not required to achieve particle
trapping at the groove entrance.  Most interestingly, the out-of-plane ($x$)
component of particle velocity, $u_{\mathrm{p},x}$ at the examined $y$-$z$
cross section is non-zero everywhere in the groove, as shown by
Fig.~\ref{fig:figure-4}(c).  Consequently, the particles accumulate at a
focusing point where the total particle velocity $\mathbf{u}_\mathrm{p}$ is
non-vanishing and, thus, they are continuously transported in and out of the
peak region.  The 3D streamlines of the particle velocity field can be seen in
Fig.~S3. It is worth noting that $\mathbf{u}_\mathrm{p}$ vanishes at the center
of the flow recirculation at the $y$-$z$ cross section $x=0$, since
$u_{\mathrm{p},x}=0$ due to symmetry.  Counterintuitively, in the examined
system the particle concentration peak is not achieved at that position despite
$\mathbf{u}_\mathrm{p}=0$ and $\mathbf{\nabla} \cdot \mathbf{u}_\mathrm{p}<0$.
Note that the latter relation is a necessary condition for particle focusing
(see derivation in SI), whereas the former is neither necessary nor sufficient
for focusing to occur.

Our physical interpretation of the trapping mechanism is validated also by the fact that
the simulation predicts a location of the peak in the particle concentration profile
along the channel depth (Fig.\ref{fig:figure-2}) at $y/H=0.052$, which agrees very well
with the peak location observed in the experiments (i.e., $y/H=0.051\pm0.003$).
Furthermore, the trapping efficiency, $A=0.69$, in the
simulations compares well with the one 
calculated from experiments ($A=0.79\pm0.03$).

\begin{figure}[t!] \centering
\includegraphics[width=0.93\columnwidth]{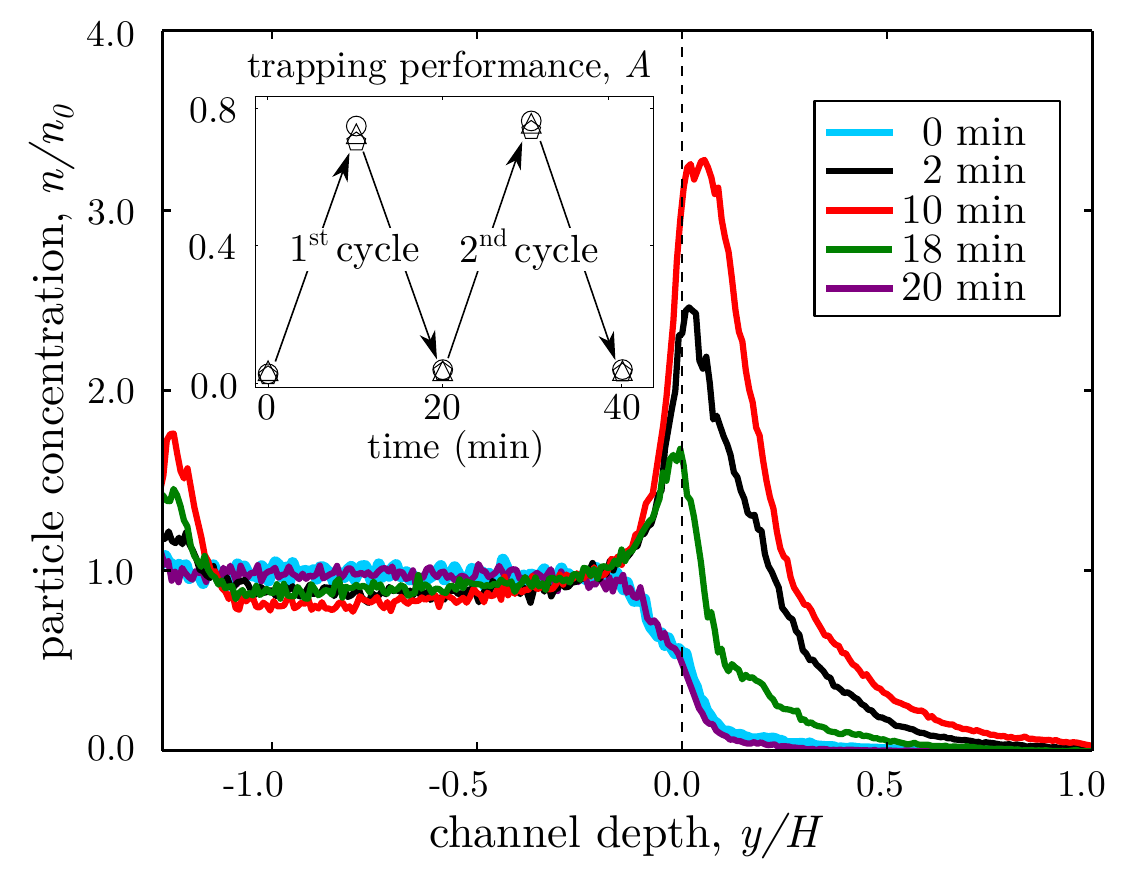}%
\caption{Reversibility of the particle trapping phenomenon. Time evolution
of particle concentration profiles as the colloids are delivered and extracted
from the groove by alternating the outer flow between the LiCl solutions
at $c_\HH=10$~mM and $c_\LL=0.1$~mM.
Inset: trapping performance of three neighboring grooves during
two delivery/extraction cycles.  \label{fig:figure-5}} \end{figure}
Upon removal of the salt gradient, the trapping mechanism ceases
and the particles can freely diffuse out of the grooves. Such an effect
allows one to control the delivery and extraction of particles into and from the grooves
by simply adjusting the salinity contrast between the inner and outer flows.
This capability is confirmed by the experimental
results, shown in Fig.~\ref{fig:figure-5}, where
the outer flow is alternated between two LiCl solutions of $c_\LL=0.1$~mM and $c_\HH=10$~mM.
As the flow configuration is cycled
between iso-osmotic ($t=0$, $20$, $40$~min) and salt gradient
($t=10$, $30$~min) conditions, the colloid concentration profile and the groove trapping performance
change accordingly over time and return to the initial values at the end of
each cycle. 

To conclude, this letter demonstrates a new mechanism for reversible trapping
of sub-micron particles in dead-end geometries under steady-state gradients. 
The key ingredients, enabling the fast and steady
accumulation of colloids, 
are the diffusiophoresis migration of particles along the groove depth direction and the flow
recirculation region below the groove entrance.  Also, the non-vanishing
particle velocity in the focusing region prevents sub-micron particles from
clustering and permanently clogging the grooves.  As a result, the trapping
phenomenon is fully reversible and particles can be cyclically trapped and
released.
We envisage that the proposed mechanism can unlock new opportunities for
the exploitation of DP transport in soft matter and living systems
for drug delivery, synthetic biology and on-chip diagnostics applications.
For example, time-controlled sequential delivery of multiple drugs
within dead-end pores could be achieved by switching between flow streams laden
with particles having different payloads. On-chip pre-concentration of extracellular 
microvesicles and cells, followed by sample release and off-chip 
downstream analysis (e.g. flow cytometry, SEM/TEM) could also be explored.

\begin{acknowledgments}
This research was supported by the EPSRC (EP/S013865/1 and EP/M027341/1)
	and the Santander Mobility Grant awarded to NS. 
We thank R. Fulcrand for help with the microdevice manufacturing.
\end{acknowledgments}


%

\end{document}


\title{Reversible Trapping of Colloids in Microgrooved Channels\\ via 
Diffusiophoresis under Steady-State Solute Gradients}

\author{Naval Singh}
\author{Goran T. Vladisavljevi\'{c}}
\affiliation{
	Department of Chemical Engineering, Loughborough University, Loughborough, LE11 3TU, United Kingdom}
\author{Fran\c{c}ois Nadal}
\affiliation{
Wolfson  School  of  Mechanical,  Electrical  and  Manufacturing  Engineering, 
Loughborough  University, Loughborough,  LE11  3TU,  United  Kingdom}
\author{C\'{e}cile Cottin-Bizonne}
\author{Christophe Pirat}
\affiliation{
Institut Lumi\`{e}re Mati\`{e}re, UMR5306 Universit\'{e} Claude Bernard Lyon 1
- CNRS, Universit\'{e} de Lyon, Villeurbanne Cedex, 69622, France} 
\author{Guido Bolognesi}
\email{g.bolognesi@lboro.ac.uk}
\homepage{www.particlemicrofluidics.com}
\affiliation{
	Department of Chemical Engineering, Loughborough University, Loughborough, LE11 3TU, United Kingdom}

\date{\today}

\maketitle

\section{Supplementary Information}
\subsection{Device Fabrication and Operation}
\begin{figure}[h!] \centering
	\includegraphics[width=0.5\columnwidth]{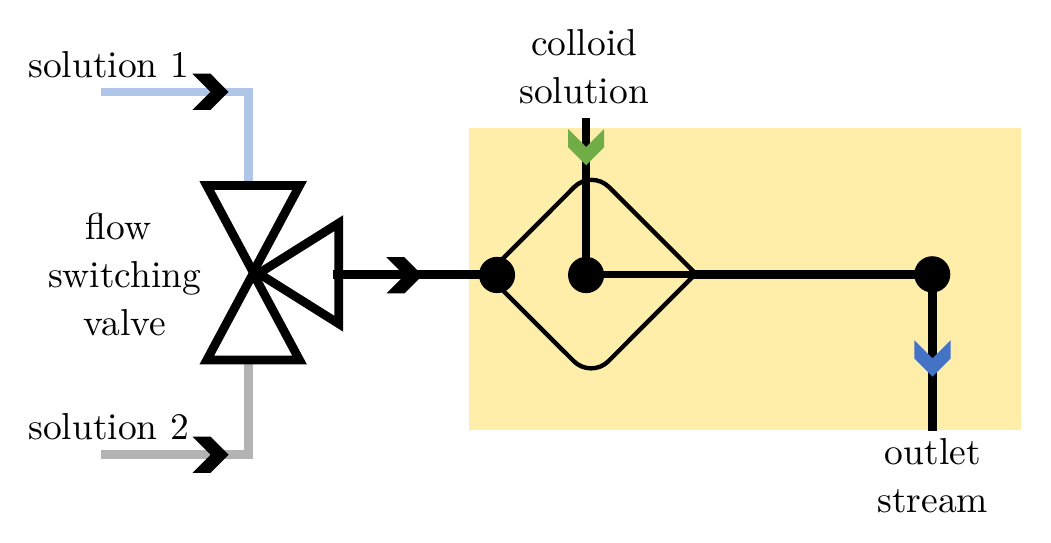}%
\caption{Device flow configuration. The $\Psi$-junction device is fed by a
	colloidal solution for the inner flow and either solution 1 or 2 for
	the outer flow. A flow switch valve is used to switch the outer flow
	between solutions 1 and 2.  \label{fig:flow_setup}} \end{figure}

The $\Psi$-shaped microchannel is fabricated using standard photo- and soft-lithography 
processes, such as PDMS replica molding from a negative photoresist SU8-2050 (MicroChem Crop., Newton,
USA) master on a silicon surface and microfluidic stickers
soft imprint lithography technique~\cite{Bartolo2008a}. The photo curable
thiolene based resin NOA 81 (Norland optical adhesive, USA) patterned mask is
prepared from a PDMS (Techsil, UK) mold and partially cured by UV lamp
(intensity 3 mW/cm$^2$). The adhesive-imprinted channel is then sealed by UV
curing against a microgrooved silicon substrate fabricated by deep reactive ion
etching. The dimensions of the channel are width, $w=400$~$\mu$m and
depth, $h=57$~$\mu$m. The dimensions of the microgrooves are thickness, $T=8$~$\mu$m,
groove-pitch, $L = 32$~$\mu$m and height, $H = 45$~$\mu$m. 
The total length of the grooved wall is $4$~cm, thereby including 1250~grooves in total.
Albeit the adhesion is strong, the device is further heated at 80$^\circ$C for two hours using
a hot plate to handle high pressure flows.
The inner solution with colloidal particles is supplied by a syringe pump at a flow rate of $12.5$~$\mu$L/min.
The outer flow can be switched between two solutions, as shown in Fig.~\ref{fig:flow_setup}, by means
of a flow switching valve. Both solutions are injected by means of a syringe pump at a flow rate of
$12.5$~$\mu$L/min.

\subsection{Image Acquisition and Analysis}
\begin{figure}[t!] \centering
	\includegraphics[width=0.5\columnwidth]{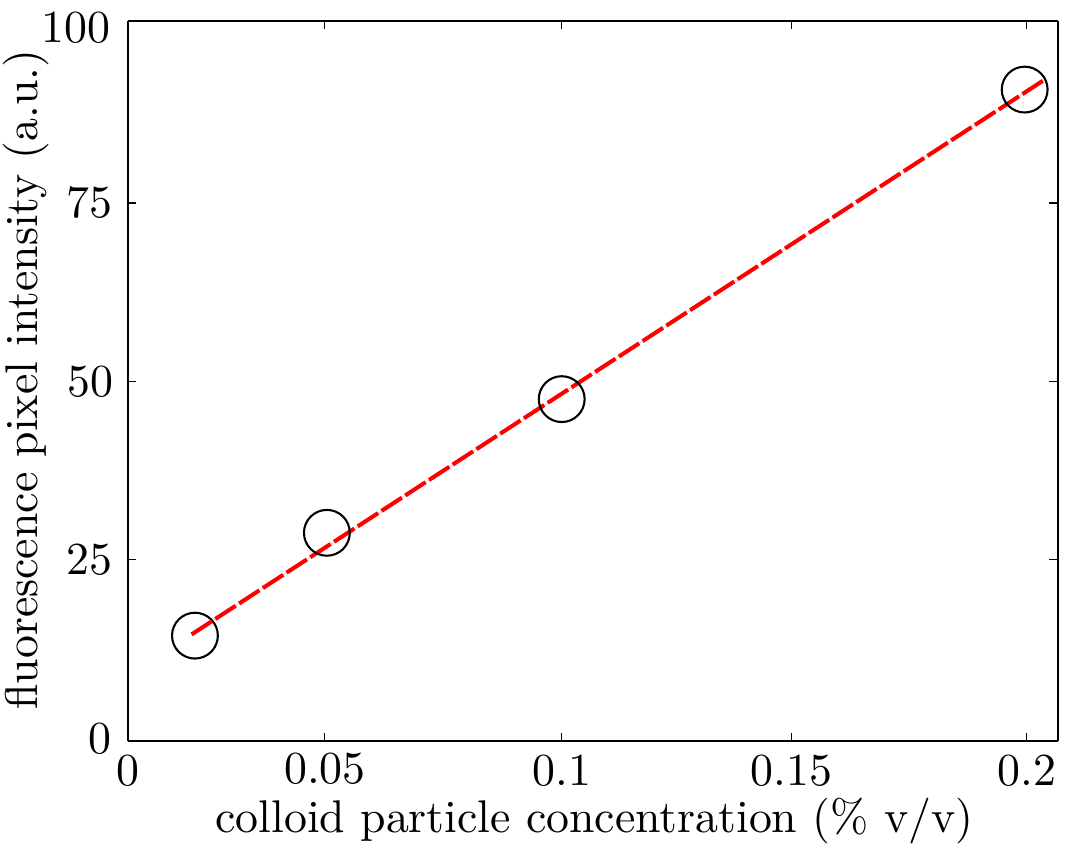}%
\caption{Confocal microscope calibration curve: particle concentration VS fluorescence intensity.\label{fig:calibration}}
\end{figure}

Fluorescent colloidal particles concentration recording and imaging are
captured using laser scanning confocal microscopy system (Leica TCS SP5, Leica
Microsystems) with a 63X Leica water immersion objective (N.A. 1.2, HCX PL APO
CS). The excitation wavelength of the dicroic beam splitter is 543~nm and HyD
hybrid detector is set between 590~nm and 799~nm. For recording each z-scan, 
the focus is shifted to the top wall of the microfluidic channel and
whole channel is scanned until the bottom of the microgrooves. The colloidal
particle distribution is captured as z~stack of $512\times512$ pixels 16-bit TIFF images,
acquired at a constant step size ($\Delta$z) of 378~nm. A full z-scan acquisition
takes approximately 60 seconds. 
The fluorescence intensity ($I$) from samples of varying particle concentrations $n$ is also recorded and
a calibration curve $n$ vs $I$ is determined (Fig.~\ref{fig:calibration}). The curve is fitted to a linear function, $I=\alpha n + I_0$,
with $\alpha$ and $I_0$ the fitting parameters.

\subsection{Evaluation of the Diffusiophoresis Coefficient}
Particle diffusiophoresis coefficient is calculated through the formula provided
by Prieve and co-workers~\cite{prieve1984motion}
\begin{equation} \Gamma_{\DP} = \frac{\varepsilon}{2\eta} \left(
\frac{k_b T}{Z e} \right)^2 \left[\bar{u}_0 + \bar{u}_1\:\lambda + \mathcal{O}(\lambda^2)
\right] \label{eq:gDP} \end{equation}
with $\varepsilon$ the absolute permittivity of the medium, $\eta$ the medium
viscosity, $k_b T$ the thermal energy, $Z$ the ion valence, $e$ the elementary
charge and $\lambda=(\kappa a)^{-1}$ is the ratio between the Debye length
$\kappa^{-1}$ and the particle radius $a$.  The Debye length $\kappa^{-1}$ is
given by
\begin{equation} \kappa^{-1} = \left( \frac{ 2 (Z e)^2 C_{\infty} } {\varepsilon k_b
T} \right)^{1/2} \label{eq:kappa} \end{equation}
with $C_{\infty}$ the density number of the solute in the medium.  The
$0^\mathrm{th}$ and 1$^\mathrm{st}$ order terms in the expansion of
$\Gamma_\DP$ for small values of $\lambda$ are given by
\begin{equation} \bar{u}_0 = 2 \beta \bar{\zeta} - 4 \ln (1-\gamma^2)
\label{eq:DP_lam_u0} \end{equation}
\begin{equation} \bar{u}_1 = F_0 + \beta F_1 + \mathrm{Pe}_D \left[ F_2 + \beta
\left( F_3 + F_5 \right) + \beta^2 F_4 \right] \label{eq:DP_lam_u0}
\end{equation}
with $\bar{\zeta} = \frac{Z e \zeta}{k_b T}$ the adimensionalised zeta
potential, $\gamma = \tanh(\bar{\zeta}/4)$, $\beta = \frac{D_+ - D_-}{D_+ +
D_-}$ and $D_+$ and $D_-$ the diffusivities of cations and anions, respectively.  
The adimensional number $Pe_D$ is given by
\begin{equation} Pe_D=\frac{\varepsilon}{2 \eta D_\mathrm{s}}\left(
\frac{k_b T}{Z e} \right)^2 \label{eq:DP_lam_Pe_D} \end{equation}
with $D_\mathrm{s}=\frac{2D_+ D_-}{D_+ + D_-}$ the salt diffusivity.  The functions
$F_n(\bar{\zeta})$ are calculated by linearly interpolating the numerical
evaluations of $F_n$ functions provided in~\cite{prieve1984motion}.  Dynamic light
scattering and electrophoretic light scattering measurements, performed with an Anton Paar
Litesizer 50 instrument, provides an average particle 
in diameter $215\pm6$~nm 
and
zeta potential of $-82\pm1$~mV.  By using $D_+=1.026\times10^{-9}$~m$^2$/s and
$D_-=1.964\times10^{-9}$~m$^2$/s for LiCl~\cite{tanaka1987measurements,mills1957self} and
$C_\infty\simeq 3$~mM, Eq.(\ref{eq:gDP}) gives
$\Gamma_\DP=83$~$\mu$m$^2$/s when second and higher order terms in
$\lambda$ are neglected. Under these conditions, the value of $\lambda$ is
$52.7\times10^{-3}$.

\subsection{Numerical Simulations}

The 3D computational domain consists in a rectangular channel with a single
groove located at a distance $z_g$ from the channel inlet.  The following set
of dimensionless equations for the hydrodynamic velocity $\mathbf{u}$, pressure
$p$, salt concentration $c$ and particle concentration $n$ are solved in Comsol
Multiphysics$^\mathrm{\textregistered}$
\begin{equation} \mathrm{Re} \: \mathbf{u}\cdot\mathbf{\nabla}{\mathbf{u}} =
-\mathbf{\nabla}{p} + \nabla^2{\mathbf{u}} \label{eq:NS} \end{equation}
\begin{equation} \nabla\cdot \mathbf{u}=0 \label{eq:a_continuity}
\end{equation}
\begin{equation} \mathrm{Pe_c} \: \mathbf{u} \cdot \mathbf{\nabla}c =  \nabla^2
c \label{eq:a_salt} \end{equation}
\begin{equation} \mathrm{Pe_n}\:	\nabla \cdot [ (\mathbf{u} +
\mathbf{u}_\DP)\: n ] = \nabla^2 n \qquad \text{with
}\mathbf{u}_\DP= \xi_\DP \frac{\mathbf{\nabla} c}{c}
\label{eq:a_particle} \end{equation}
with all quantities rescaled according to the following relations
\begin{equation} u\propto U_0   \qquad x,y,z\propto w  \qquad p\propto
\frac{\eta U_0}{w} \qquad c\propto c_{H}  \qquad n\propto n_0
\label{eq:scale_var} \end{equation}
where $U_0$ is the average hydrodynamic velocity along the flow direction, $w$
is the actual (dimensional) channel width, $\eta$ is the viscosity of both
outer and inner solutions, $c_H$ is the solute concentration of the outer
solution, $n_0$ is the particle concentration of the inner solution.  At the
channel inlet, the boundary condition for the velocity field is
$\mathbf{u}=\mathbf{u}_\mathrm{inlet}$, with $\mathbf{u}_\mathrm{inlet}$ the
fully developed velocity field at a cross section of the rectangular channel
perpendicular to the flow direction and with average velocity equal 1.  The
boundary conditions at channel inlet for the salt and concentration fields are
$c=1$ and $n=0$ for the outer flow region and $c=c_L/c_H$ and $n=1$ for the
inner flow region.  At the channel outlet, the zero normal gradient boundary
condition for the pressure, salt and particle concentrations are imposed.  At
the remaining walls, the slip boundary condition $\mathbf{u} = -\xi_{\DO}
\frac{\mathbf{\nabla} c}{c}$ is applied together with the zero flux condition
for the salt and particle concentration fields.  The channel outlet was located
at 5 times the channel depth from the groove to ensure that the boundary
conditions at the channel outlet do not affect the fields near the groove.  The
dimensionless numbers, governing the examined system, are defined as follows
\begin{equation} \mathrm{Re}=\frac{\rho \: U_0 \: w}{\eta} \qquad
\mathrm{Pe_c}=\frac{w \: U_0}{D_\mathrm{s}} \qquad \mathrm{Pe_n}=\frac{w \:U_0}{D_\mathrm{p}}
\qquad \mathrm{\xi_\DP}=\frac{\Gamma_\DP}{w\: U_0} \qquad
\mathrm{\xi_\DO}= \frac{\Gamma_\DO}{w \: U_0}
\label{eq:adimensional_group} \end{equation}
where $\rho$ is the density of the inner and outer solutions, $D_\mathrm{p}$ is the
particle diffusivity, calculated as $k_b T/ (6 \pi \eta a) $, and
$\Gamma_\DO$ is the diffusioosmosis coefficient of the channel walls.
Despite in the experiments the channel walls are made of different materials
(namely, silicon for the grooved substrate and optical adhesive glue for the
remaining channel walls), a single value of $\Gamma_\DO$ is used for
all channel walls in the simulations to take advantage of the symmetries of the
problem, hence limiting the computational cost. The good agreement between the
numerical simulations and experimental results suggests that this assumption is
acceptable.

Since the hydrodynamic velocity and salt concentration fields are coupled by the
boundary condition for the diffusioosmosis slip velocity, both fields should be
solved simultaneously. The resulting finite element computation on a 3D domain would
involve a large number of degrees of freedom that would likely require the use
of an expensive work station with large memory. Conversely, we can reformulate the problem
in order to solve it on a standard PC with 2.7 GHz dual-core processor and 16
GB memory. By using a perturbation approach, the velocity, pressure and salt concentration field are
expressed as a power expansion of the parameter $\xi_\DO$ as follows

\begin{equation}
	\mathbf{u}=\mathbf{u}_0+\xi_\DO\:\mathbf{u}_1+\mathcal{O}(\xi_\DO^2) 
	\label{eq:pert_exp_vel}
\end{equation}
\begin{equation}
	p=p_0+\xi_\DO\:p_1+\mathcal{O}(\xi_{DO}^2) 
	\label{eq:pert_exp_pres}
\end{equation}
\begin{equation}
	c=c_0+\xi_\DO\:c_1+ \mathcal{O}(\xi_{DO}^2) 
	\label{eq:pert_exp_salt}
\end{equation}
The 0$^{th}$ and 1$^{st}$ order terms of the fields are obtained by solving the following equations
\begin{equation}
	\mathrm{Re} \: \mathbf{u}_0\cdot\mathbf{\nabla}{\mathbf{u}_0} = -\mathbf{\nabla}{p}_0 + \nabla^2{\mathbf{u}_0}
	\label{eq:NS_ord_0}
\end{equation}
\begin{equation}
	\nabla\cdot \mathbf{u_0}=0
	\label{eq:a_continuity_ord_0}
\end{equation}
\begin{equation}
	\mathrm{Pe_c} \: \mathbf{u_0} \cdot \mathbf{\nabla}c_0 =  \nabla^2 c_0
	\label{eq:a_salt_ord_0}
\end{equation}
\begin{equation}
	\mathrm{Re} \: \left( \mathbf{u}_1\cdot\mathbf{\nabla}{\mathbf{u}_0}+ \mathbf{u}_0\cdot\mathbf{\nabla}{\mathbf{u}_1} \right) = -\mathbf{\nabla}{p}_1 + \nabla^2{\mathbf{u}_1}
	\label{eq:NS_ord_1}
\end{equation}
\begin{equation}
	\nabla\cdot \mathbf{u_1}=0
	\label{eq:a_continuity_ord_1}
\end{equation}
\begin{equation}
	\mathrm{Pe_c} \: \left( \mathbf{u_0} \cdot \mathbf{\nabla}c_1 +  \mathbf{u_1} \cdot \mathbf{\nabla}c_0 \right) =  \nabla^2 c_1
	\label{eq:a_salt_ord_1}
\end{equation}
The wall boundary conditions for the velocity fields are $u_0=0$ and
$u_1=-\mathbf{\nabla}{c_0}/c_0$. As a results, velocity and concentration fields can be now solved separately
in the following order: $u_0$, $c_0$, $u_1$ and $c_1$. The particle concentration $n$
is determined at last by solving Eq.(\ref{eq:a_particle}) with the particle
diffusiophoresis velocity expressed as 
\begin{equation}
\mathbf{u}_\DP=\xi_\DP
	\frac{ \mathbf{\nabla} \left( c_0 + \xi_\DO c_1 \right) }{  c_0 + \xi_\DO c_1} 
\end{equation}
The numerical results presented in the manuscript are obtained by using
the parameters shown in Table~\ref{tab:num_val}.  The value of $\Gamma_{DO}$
for the materials of the microchannels used in our experiment is not known, so
this parameter is adjusted in order to achieve a good match between
experimental and numerical results.  The adjusted value of
$\Gamma_{DO}=375$~$\mu$m$^2$/s, which corresponds to $\xi_{\DO}=4.5\: \xi_\DP$,
is of the same order of measured DO coefficients 
for silicon substrates under similar experimental conditions~\cite{lee2014osmotic}.
\begin{table}[h!] \begin{center} \begin{tabular}{ |c |c | } \hline
	\textbf{Parameter} & \textbf{Value}\\ \hline \hline $\rho$ &
	$10^3\:$kg/m$^3$  \\ \hline $\eta$ & $0.9\times10^{-3}\:$Pa$\:$s \\
	\hline $w$ & $400\:\mu$m \\
	\hline $h$ & $57\:\mu$m \\
	\hline $T$ & $8\:\mu$m \\
	\hline $H$ & $45\:\mu$m \\
	\hline $z_g$ & $4$~mm \\ \hline $U_0$ & 18.28 mm/s \\ \hline $c_L$&
	0.1 mM\\ \hline $c_H$& 10 mM\\ \hline $2a$& 215 nm\\ \hline $\Gamma_\DP$ &
	$83.2$ $\mu$m$^2$/s \\ \hline $\Gamma_\DO$ &
	$375$ $\mu$m$^2$/s \\ \hline Re & 8.12 \\ \hline Pe$_c$ &
	5423  \\ \hline Pe$_n$ &3.24$\times$10$^6$  \\ \hline $\xi_\DO$
& $51.2\times10^{-6}$  \\ \hline $\xi_\DP$ & $11.4\times10^{-6}$  \\
\hline \end{tabular} \end{center} \caption{Simulation Parameters}
\label{tab:num_val} \end{table}
\begin{figure}[t!] \centering
	\includegraphics[width=0.9\textwidth]{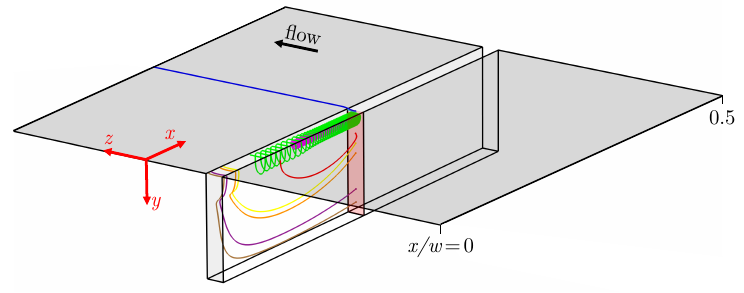}%
	\caption{Groove at $4\:$mm from the junction and 3D streamlines of the particle velocity,
$\mathbf{u}_\mathrm{p}=\mathbf{u}+\mathbf{u}_\DP$, starting from points located at $x/w=0.25$ (red shaded region) and varying depths down the groove.
\label{fig:streamlines}} \end{figure}

Fig.~\ref{fig:streamlines} shows a set of 3D streamlines of the particle 
velocity field, $\mathbf{u}_\mathrm{p}=\mathbf{u}+\mathbf{u}_\DP$,
starting at the $y$-$z$ cross section, $x=0.25$, and at varying depths down the
groove. Particles close to the groove entrance can either escape from the groove (blue line)
or be caught within the flow recirculation region (green and magenta lines). 
Particles further down the groove are transported towards the groove entrance due to diffusioosmosis
(remaining lines).
It is worth noting that these streamlines, calculated by integrating the
field $\mathbf{u}_\mathrm{p}$, do not correspond to the actual particle
trajectories since the Brownian diffusivity allows particle to move across adjacent flow
streamlines.

\subsection{Effect of Diffusioosmosis}
\begin{figure}[b!] \centering
\includegraphics[width=0.9\textwidth]{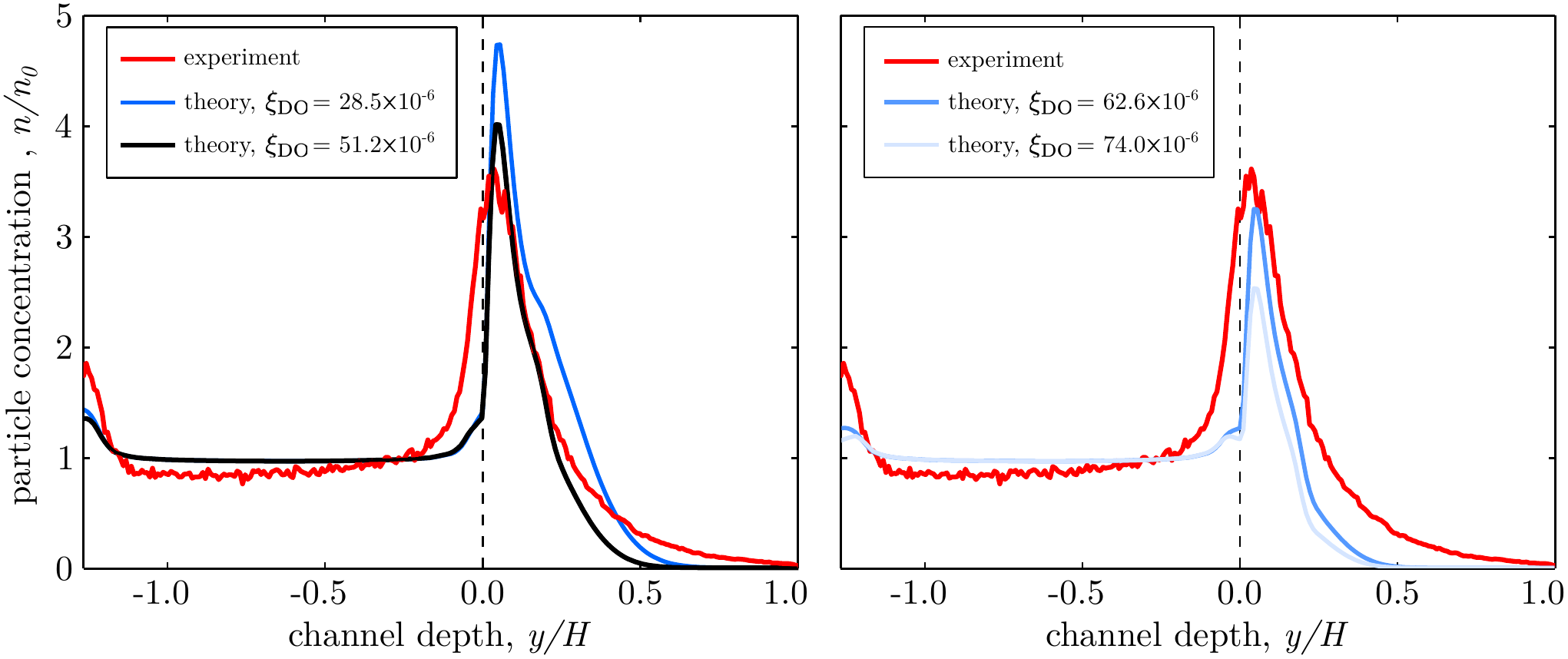}%
\caption{Numerical predictions for the particle concentration profiles along
the channel depth ($y$ axis) for varying DO coefficient values. For sake
of comparison, the experimental steady-state profile in Fig.2 of the main manuscript
is also shown (red curve).  \label{fig:DO_effect}}.  \end{figure}

To clarify the effect of diffusioosmosis (DO) on particle trapping, 
numerical simulations are performed with different values of the normalized DO
coefficient ranging from $\xi_\textrm{DO}=
2.5\:\xi_\textrm{DP}$
to $\xi_\textrm{DO}=6.5\:\xi_\textrm{DP}$, namely from
$28.5\times10^{-6}$ to $74.0\times10^{-6}$. The corresponding
$x$-/$y$-averaged particle concentration profiles along the channel depth (y axis)
are shown in~Fig.\ref{fig:DO_effect}.
All concentration profiles have a similar shape characterized by a peak
located near the entrance of the groove ($y>0$). 
Table~\ref{tab:peaks} displays the
normalized values of particle concentration peak intensity, $n_\textrm{max}/n_0$,
peak position along channel depth, $y_p/H$, and peak width, $w_p/H$ – where $w_p$
is the full width at half maximum – for the examined $\xi_\textrm{DO}$ values.
It can be concluded that, as the value of $\xi_\mathrm{DO}$ and, hence, the strength of the
DO-driven flow increases, 
the intensity and width of the concentration peak
decrease. 
The narrowing of the peak can be explained by the fact that
the peak width is governed by the depth of the recirculation region
within the groove, where particle trapping occurs. 
Since higher DO-driven flows directed from the groove towards
the channel, lead to shallower flow recirculation regions,
the width of the particle concentration peak also decreases.
This interpretation is confirmed by the data in Table~\ref{tab:peaks}
showing how the depth of the 
recirculation region, $d_\textrm{rec}$, decreases for increasing DO coefficient values.
Interestingly, the center of the recirculation region is not affected by
the DO flow, therefore the location of the concentration peak along the
channel depth, $y_p$, remains unchanged for varying DO coefficients.
As shown in~Fig.\ref{fig:DO_effect},
a good match between experimental and theoretical results is
achieved for $\xi_\textrm{DO}=4.5\:\xi_\textrm{DP}=51.2\times10^{-6}$.

\begin{table}[h!] 
\caption{Main properties of the particle
concentration peak for varying values of the DO coefficient: 
$n_\mathrm{max}$, peak intensity;
$y_p$, peak position along $y$ (depth) direction;
$w_p$, peak full width at half maximum; 
$d_\mathrm{rec}$, size of the flow recirculation region along $y$ (depth)
direction.} \label{tab:peaks} \begin{center} \begin{tabular}{|c| c| c| c| c| }
\hline $\xi_\textrm{DO}\times10^6$  & 28.5 & 51.2 & 62.6 & 74.0  \\ \hline
\hline $n_{\max}/n_0$ & 4.75 & 4.07 & 3.29 & 2.56  \\ \hline $y_p/H$  & 0.050 &
0.051 & 0.050 & 0.050   \\ \hline $w_p/H$  & 0.1081 & 0.0971 & 0.0881 & 0.0781
\\ \hline $d_\textrm{rec}/H$ &  0.1150 & 0.0990 & 0.0975 & 0.0965  \\ \hline
\end{tabular} \end{center}  \end{table}

\begin{figure}[b!] \centering
	\includegraphics[width=0.7\textwidth]{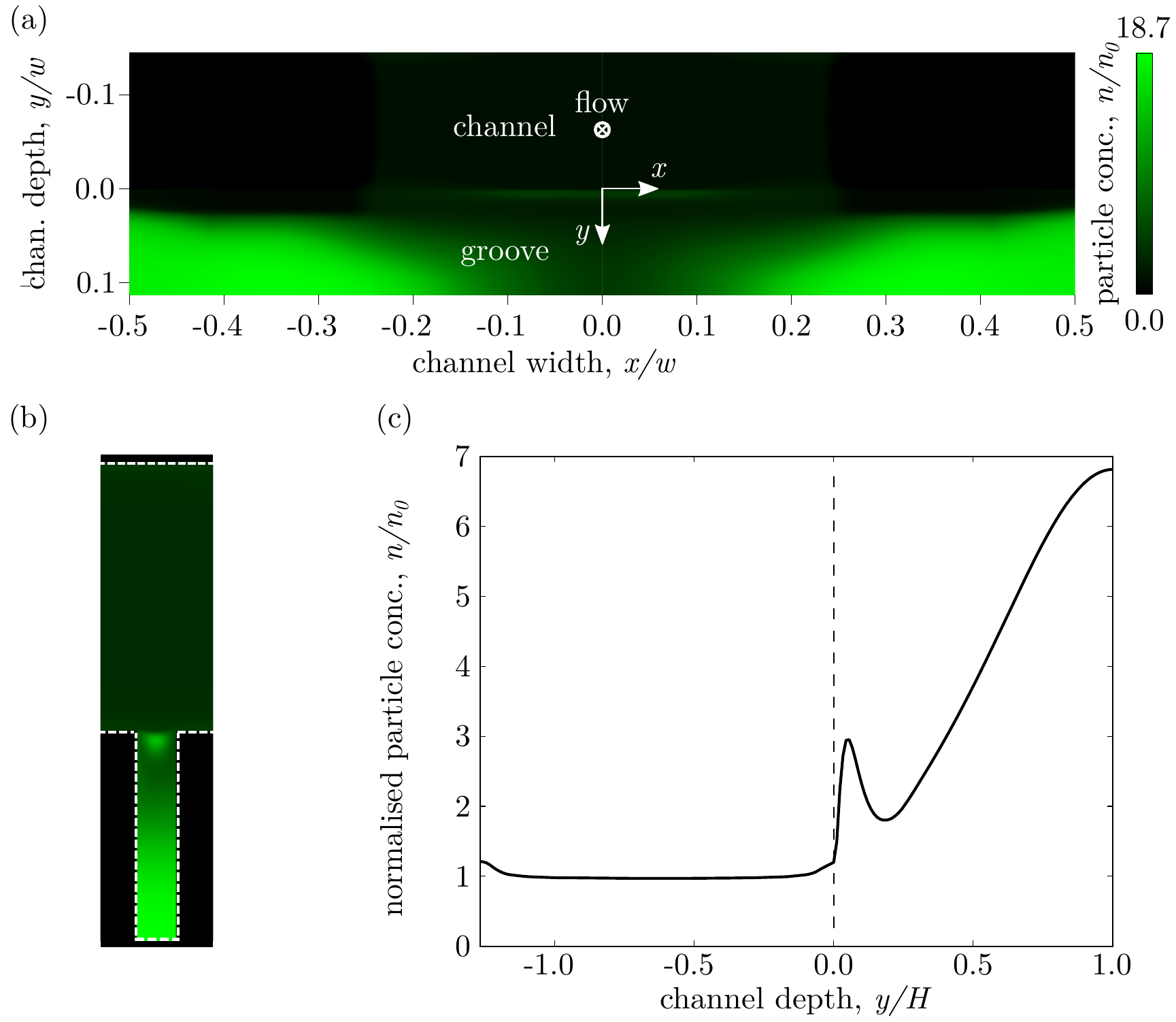}%
	\caption{Simulated particle distribution without DO (i.e.,
	$\xi_\DO=0$). (a) Particle concentration at a $x$-$y$ cross
	section, perpendicular to the flow direction $z$, at $4\:$mm from the
	junction. The particle concentration is averaged over the
        groove thickness $T$ along the $z$ axis. (b) Side view of particle distribution on a $z$-$y$ plane.
	The particle concentration is averaged over the channel width direction $x$.
	The dashed white lines represent the channel boundaries and groove
edges.  (c) $x$-/$y$-averaged particle concentration profile along the channel
depth ($y$ axis).  \label{fig:figure-sim-noDO}} \end{figure}

To understand further the role played by diffusioosmosis (DO),
numerical simulations are performed also in the absence of DO effects (i.e.
$\xi_\DO$=0).  Fig.~\ref{fig:figure-sim-noDO}(a) shows the particle
concentration at a $x$-$y$ cross section perpendicular to the flow direction
whereas Fig.~\ref{fig:figure-sim-noDO}(a) show a side view of the particle
distribution on a $z$-$y$ plane under these conditions. The corresponding
particle concentration profile along the channel depth is shown in
Fig.~\ref{fig:figure-sim-noDO}(c). 
Again, a particle concentration peak located 
just below the groove entrance and at 
the center of the flow recirculation region ($y_p/H=0.050$) can be observed,
thereby demonstrating that the DO is not required for the discussed particle trapping 
mechanism to occur. Furthermore, in the absence of a DO flow from the groove towards the 
channel, particles can now also accumulate at the bottom of the groove as shown in
Fig.~\ref{fig:figure-sim-noDO}. 

%

\subsection{Condition for particle focusing}
A focusing point corresponds to a local maximum of the particle concentration field, $n$. At this point
of local maximum, the following conditions must be verified
\begin{equation}
\nabla n  = 0 
\label{eq:n_peak_cond1}
\end{equation}
\begin{equation}
\mathrm{Tr}(\mathcal{H}) <0
\label{eq:n_peak_cond2}
\end{equation}
\begin{equation}
\mathrm{det}(\mathcal{H}) >0
\label{eq:n_peak_cond3}
\end{equation}
with $\mathcal{H}$ the Hessian matrix of the field $n(x,y,z)$
\begin{equation}
	\mathcal{H}=
	\left(
	\begin{array}{cc}
		\frac{\partial^2 n }{\partial x^2} & \frac{\partial^2 n}{\partial x \partial y} \\
		\frac{\partial^2 n}{\partial x \partial y}  &	\frac{\partial^2 n }{\partial y^2}  \\
	\end{array}
	\right)
	\label{eq:n_hes}
\end{equation}
Eq.\eqref{eq:n_peak_cond2} and Eq.\eqref{eq:n_peak_cond3} can hence be rewritten as
\begin{equation}
\nabla^2 n <0
\label{eq:n_peak_cond2b}
\end{equation}
\begin{equation}
\frac{\partial^2 n}{\partial x^2} \cdot \frac{\partial^2 n}{\partial y^2} - 2 
	\frac{\partial^2 n}{\partial x \partial y}> 0
\label{eq:n_peak_cond3b}
\end{equation}
From Eq.\eqref{eq:a_particle}, it follows
\begin{equation}
	n\: \nabla \cdot  \mathbf{u}  +
	\mathbf{u}\cdot \nabla n +
	n\: \nabla \cdot  \mathbf{u}_{\DP}  +
	\mathbf{u}_{\DP}\cdot \nabla n 
	= \frac{1}{\mathrm{Pe_n}} \nabla^2 n
	\label{eq:particle_1}
\end{equation}
Since $\nabla \cdot \mathbf{u}=0$ and, at the local peak, $\nabla n =0$,
it follows that (for $n\ne$0) 
\begin{equation}
	\nabla \cdot \mathbf{u_{\DP}}  = \frac{1}{\mathrm{Pe_n}} \frac{\nabla^2 n}{n} < 0
	\label{eq:div_v_dp_cond}
\end{equation}
It can be concluded that $\nabla \cdot \mathbf{u}_\mathrm{\DP}<0$ is a
necessary condition for particle focusing to occur.  By denoting the particle
velocity as $\mathbf{u}_\mathrm{p}=\mathbf{u}+\mathbf{u}_{\DP}$, the
same condition can also be expressed as $\nabla \cdot \mathbf{u}_\mathrm{p}<0$.

Interestingly, Eq.\eqref{eq:div_v_dp_cond} can also be re-written in term of a scalar
product between the hydrodynamic velocity field $\mathbf{u}$ and the diffusiophoresis
velocity field $\mathbf{u}_{\DP}$.  Indeed, the divergence of $\mathbf{u}_{\DP}$ can be
re-written as
\begin{equation} \nabla \cdot \mathbf{u}_{\DP} = \xi_{\DP} \nabla \cdot \left(
\frac{\nabla c}{c}  \right) = \frac{\xi_{\DP}}{c} \left( \nabla^2 c -
\frac{\nabla c \cdot \nabla c}{c} \right) \label{eq:div_v_dp} \end{equation}
By replacing the expression for the term $\nabla^2 c$ from Eq.\eqref{eq:a_salt}, it follows
\begin{equation} \nabla \cdot \mathbf{u}_{\DP} = \frac{\xi_{\DP}}{c}
\left(\mathrm{Pe_c} \: \mathbf{u} \cdot \mathbf{\nabla}c - \frac{\nabla c \cdot
\nabla c}{c} \right) = \mathrm{Pe_c} \: \mathbf{u} \cdot \mathbf{u}_{\DP}
- \frac{1}{\xi_{\DP}} \mathbf{u}_{\DP} \cdot \mathbf{u}_{\DP} 
\label{eq:div_v_dp_1} \end{equation}
The necessary condition in Eq.\eqref{eq:div_v_dp_cond} can hence be re-written as 
\begin{equation} 
\mathbf{u} \cdot \mathbf{u}_{\DP} =
\frac{\mathbf{u}_{\DP} \cdot \mathbf{u}_{\DP} }
{\xi_{\DP}\:\mathrm{Pe_c}}=\frac{D_\mathrm{s}}{\Gamma_{\DP}}
\mathbf{u}_{\DP} \cdot \mathbf{u}_{\DP} 
\label{eq:div_v_cond_1} \end{equation}
By denoting $\mathbf{\hat{u}}_{\DP}$ as the unit vector with the direction of the DP velocity,
Eq.\eqref{eq:div_v_cond_1} can be re-expressed as (for $|\mathbf{u}_{\DP}|\ne0$)
\begin{equation} 
\mathbf{u} \cdot \mathbf{\hat{u}}_{\DP} <
\frac{D_\mathrm{s}}{\Gamma_{\DP}} |\mathbf{u}_{\DP}|
\label{eq:div_v_cond_2} \end{equation}
Therefore, at a local peak of the particle concentration field,
the component of the hydrodynamic field $\mathbf{u}$ in the
direction of the DP velocity $\mathbf{u}_{\DP}$ must not exceed
$\frac{D_\mathrm{s}}{\Gamma_{\DP}}$ times the intensity of the DP velocity.  

\clearpage
\newpage
\bibliography{main_SI}